\documentclass[sigconf]{acmart}

\acmConference[ICSE 2022]{The 44th International Conference on Software Engineering}{May 21–29, 2022}{Pittsburgh, PA, USA}

\AtBeginDocument{%
  \providecommand\BibTeX{{%
    \normalfont B\kern-0.5em{\scshape i\kern-0.25em b}\kern-0.8em\TeX}}}

\copyrightyear{2022}
\acmYear{2022}
\setcopyright{acmcopyright}\acmConference[ICSE '22]{44th International Conference on Software Engineering}{May 21--29, 2022}{Pittsburgh, PA, USA}
\acmBooktitle{44th International Conference on Software Engineering (ICSE '22), May 21--29, 2022, Pittsburgh, PA, USA}
\acmPrice{15.00}
\acmDOI{10.1145/3510003.3510214}
\acmISBN{978-1-4503-9221-1/22/05}

\usepackage{multirow}
\usepackage{balance}
\usepackage{amsmath}
\usepackage[framemethod=tikz]{mdframed}
\mdfdefinestyle{mpdframe}{
    frametitlebackgroundcolor   =black!15,
    frametitlerule              =true,
    roundcorner                 =5pt,
    middlelinewidth             =1pt,
    innermargin                 =0.2cm,
    outermargin                 =0.2cm,
    innerleftmargin             =0.2cm,
    innerrightmargin            =0.2cm,
    innertopmargin              =0.2cm,
    innerbottommargin           =0.2cm
}

\graphicspath{ {./images/} }

\begin{document}

\title[Detecting False Alarms from Automatic Static Analysis Tools: How Far are We?]{Detecting False Alarms from Automatic Static Analysis Tools: \\How Far are We?}

\author{Hong Jin Kang}
\affiliation{%
  \institution{Singapore Management University}
  \city{Singapore}
  \country{Singapore}}
\email{hjkang.2018@phdcs.smu.edu.sg}

\author{Khai Loong Aw}
\affiliation{%
  \institution{Singapore Management University}
  \city{Singapore}
  \country{Singapore}}
\email{klaw.2020@scis.smu.edu.sg}

\author{David Lo}
\affiliation{%
  \institution{Singapore Management University}
  \city{Singapore}
  \country{Singapore}}
\email{davidlo@smu.edu.sg}

\begin{abstract}
Automatic static analysis tools (ASATs), such as Findbugs, have a high false alarm rate. 
The large number of false alarms produced poses a barrier to adoption. 
Researchers have proposed the use of machine learning to prune false alarms and present only actionable warnings to developers. 
The state-of-the-art study has identified a set of ``Golden Features'' based on  metrics computed over the characteristics and history of the file, code, and warning. 
Recent studies show that machine learning using these features is extremely effective 
and that they achieve almost perfect performance.

We perform a detailed analysis to better understand the strong performance of the ``Golden Features''. 
We found that 
several 
studies used an experimental procedure that results in data leakage and data duplication, which are subtle issues with significant implications. 
Firstly, the ground-truth labels have leaked into features that measure the proportion of actionable warnings in a given context. 
Secondly, many warnings in the testing dataset appear in the training dataset. 
Next, we demonstrate limitations in the warning oracle that determines the ground-truth labels, a heuristic comparing warnings in a given revision to a reference revision in the future.
We show the choice of reference revision influences the warning distribution.
Moreover, the heuristic produces labels that do not agree with human oracles.
Hence, the strong performance of these techniques previously seen is overoptimistic of their true performance if adopted in practice. 
Our results convey several lessons and provide guidelines for evaluating false alarm detectors.

\end{abstract}




\maketitle

\section{Introduction}

It has been 15 years since Findbugs~\cite{ayewah2008using} was introduced to detect bugs in Java programs.
Along with other automatic static analysis tools (ASATs)~\cite{rutar2004comparison,sadowski2018lessons,distefano2019scaling}, 
FindBugs aims to detect incorrect code by matching code against bug patterns~\cite{ayewah2008using,hovemeyer2004finding}, 
for example, patterns of code that may dereference a null pointer.
Since then, many projects have adopted these tools as they help in detecting bugs at low cost.
However, these tools do not guarantee that the warnings are real bugs.
Many developers do not perceive the warnings by ASATs to be relevant
due to the high incidence of effective false alarms~\cite{johnsonwhy,vassallo2020developers,sadowski2018lessons}.
Prior work has suggested that the false positive rate may range up to 91\%.
While the overapproximation of static analysis may cause false alarms,
false alarms do not only refer to errors from analysis or overapproximation,
but include warnings that developers did not act on~\cite{johnsonwhy,sadowski2015tricorder,sadowski2018lessons}.
Developers may not act on a warning if they do not think the warning represents a bug or believe that a fix is too risky.

To address the high rate of false alarms, 
many researchers~\cite{wanggolden,heckman2011systematic} have proposed techniques to prune false alarms 
and identify actionable warnings,
which are the warnings that developers would fix. 
These approaches~\cite{hanam2014finding,heckman2008establishing,kim2007prioritizing,kremenek2004correlation,ruthruff2008predicting,yuksel2013automated,williams2005automatic,shen2011efindbugs,koc2019empirical} consider different aspects of a warning reported by Findbugs in a project, 
including factors about the source code~\cite{hanam2014finding}, repository history~\cite{williams2005automatic}, file characteristics~\cite{liang2010automatic,yuksel2013automated}, 
and historical data about fixes to Findbugs warnings~\cite{kremenek2004correlation} within the project.
Wang et al.~\cite{wanggolden} completed a systematic evaluation of the features that have been proposed in the literature 
and identified 23 ``Golden Features'', which are the most important features for detecting actionable Findbugs warnings.
Using these features, subsequent studies ~\cite{yangeasy,yangincremental,yang2021documenting} show 
that any machine learning technique, e.g. SVM, performs effectively
and that the use of a small number of training instances can train effective models.
In these studies, performances of up to 96\% Recall and 98\% Precision, and 99.5\% AUC can be achieved.
A perfect predictor has a Recall, Precision, and AUC of 100\%, suggesting that machine learning techniques using the Golden Features are almost perfect.

Although the Golden Features have been shown to perform well, we do not know why they are effective.
Therefore, in this work we seek to get a deeper understanding of the Golden Features.
We find a few issues: 
First, the ground-truth label was leaked into the features measuring the proportion of actionable warnings in a given context.
Second, warnings in the  test data were used for training.
To understand their impact, we addressed the two flaws and found that the performance of the Golden Features
declines.
Our results show that the use of the Golden Features do not substantially outperform a strawman baseline that predicts all warnings are actionable.

Next, we investigate the warning oracle used to obtain ground-truth labels when constructing the dataset.
To evaluate any proposed approach,
a large dataset should be built, where each warning is accurately labeled as either an actionable warning or a false alarm.
Many studies~\cite{wanggolden,yangeasy,yangincremental} use a heuristic, which we term the \textit{closed-warning heuristic}, as the warning oracle to determine the actionability of a warning, 
checking if the same warning is reported in a \textit{reference revision}, a revision chronologically after the \textit{testing revision}.
If the file is still present and the warning is not reported in the reference revision, then the warning is \textit{closed} and is assumed to be fixed. It is, therefore, assumed to be actionable.
Conversely, a warning that remained \textit{open} is a false alarm.
A revision made a few years after the simulated time of the experimental setting is used as the reference revision.
Prior studies~\cite{wanggolden,yangeasy,yangincremental} selected reference revisions set 2 years after the testing revision.
However, no prior work has investigated the robustness of the heuristic.

There are several desirable qualities of a warning oracle.
Firstly, it should allow the construction of a sufficiently large dataset.
Secondly, it should be reliable; the labels should be robust to minor changes in the oracle.
Thirdly, it should generate labels that human annotators and developers of projects using ASATs agree with.
An advantage of the closed-warning heuristic is that it enables the construction of a large dataset.
However, our experiments demonstrate the lack of consistency in the labels given changes in the choice of the reference revision.
This may allow different conclusions to be reached from the experiments.
Our experiments also uncover that the oracle does not always produce labels that human annotators or developers agree with.
These limitations show that alone, the heuristic do not always produce trustworthy labels.
After removing unconfirmed actionable warnings, the effectiveness of the Golden Features SVM improves,
indicating the importance of clean data.

Finally, we highlight lessons learned from our experiments.
Our results show the need to carefully design an experimental procedure to assess future approaches,
comparing them against appropriate baselines.
Our work points out open challenges in the design of a warning oracle for the construction of a benchmark. 
Based on the lessons learned, we outline several guidelines for future work.

We make the following contributions:
\begin{itemize}
  \item We analyze the reasons for the strong performance from the use of the ``Golden Features'' observed in prior studies. Contrary to prior work, we find that machine learning techniques are not almost perfect, and that there is still much room for improvement for future work in this area.
  \item We study the warning oracle, the closed-warning heuristic, that assigns labels to warnings used in previous studies. We show that the heuristic may not be sufficiently robust.
  \item We discuss the lessons learned and their implications. Importantly, we highlight the need for community effort in building an accurate benchmark and suggest that future studies compare new approaches with strawman baselines.
\end{itemize}

The rest of the paper is structured as follows.
Section \ref{sec:background} covers the background of our work.
Section  \ref{sec:study} presents the design of the study.
Section \ref{sec:golden_features} analyzes the Golden Features.
Section \ref{sec:dataset_analysis} investigates the closed-warning heuristic.
Section \ref{ref:discussion} discusses lessons learned from our study.
Section \ref{sec:related} presents related work.
Finally, Section \ref{sec:conclusion} concludes the paper.

	
\section{Background}
\label{sec:background}

\subsection{Automatic Static Analysis Tools}

Many researchers have proposed Automatic Static Analysis Tools (ASATs), such as Findbugs~\cite{ayewah2008using}, to detect possible bugs during the software development process.
Research has shown these tools are useful and succeed in detecting  bugs that developers are interested in at low cost.
Compared to program verification or software testing, these tools rely on bug patterns written by the authors of the static analysis tools, 
matching code that may be buggy.
Findbugs includes over 400 bug patterns that match a range of possible bugs, such as class casts that are impossible, 
null pointer dereferences, and incorrect synchronization.

Studies have also shown that ASATs are able to detect real bugs~\cite{thung2012extent,habib2018many}.
Indeed, static analysis tools are adopted by large companies~\cite{ayewah2010google,sadowski2018lessons,distefano2019scaling} 
and open source projects~\cite{beller_oss} to detect bugs.
Developers may run them during local development,
use them in Continuous Integration~\cite{zampetti2017open}
and during code review to detect buggy code  to catch bugs early~\cite{vassallo2020developers,tomasdottir2017and,balachandran2013reducing,panichella2015would}.
Projects may configure the tools~\cite{zampetti2017open}, for example, to
suppress false alarms by configuring a filter file to exclude specific warnings~\cite{findbugs_filter}.

Developers largely perceive ASATs to be relevant, 
and the majority of practitioners have used or heard of ASATs~\cite{tahaei2021security,vassallo2020developers,marcilio2019static}.
Still, these tools are characterized by the large amounts of false alarms that they produce,
and among other reasons, this has led to resistance in adopting them in many software projects~\cite{johnsonwhy}.

\subsection{Distinguishing between Actionable Warnings and False Alarms}

To minimize the overhead of inspecting false alarms, researchers have proposed approaches based on machine learning 
to rank or classify the warnings.
A large number of features have been designed over the past 15 years; 
for example, based on software metrics (e.g. size of the file, number of comments in the code),
source code history (e.g. number of lines of code recently added to a file), 
and characteristics and history of the warnings (e.g. the number of revisions where the warning has been opened).

Researchers have evaluated their proposed tools through datasets of warnings produced by Findbugs~\cite{hanam2014finding,heckman2008establishing,heckman2009model,ruthruff2008predicting,shen2011efindbugs,yuksel2013automated,heckman2013comparative}.
Recently, Wang et al.~\cite{wanggolden} performed a systematic analysis of the features proposed in the literature.
From 116 features, they identified 23 \textit{Golden Features},
which are the features that achieve effective performance.
The features are listed in Table \ref{tab:golden_features}.
These features include metrics such as the code-to-comments ratio~\cite{liang2010automatic}, 
and the number of lines added in the past~\cite{heckman2009model,ruthruff2008predicting}.
Of note are several features of the ``Warning combination'' feature type.
We will refer to three of these features, \textbf{warning context in method},
\textbf{warning context in file},
and
\textbf{warning context for warning type},
as the \textbf{warning context} features.
We refer to another two features, the
\textbf{defect likelihood for warning pattern},
and
\textbf{discretization of defect likelihood}
as the \textbf{defect likelihood} features.
These features are various measures of the proportion of actionable warnings within a population of warnings,
building on top of the insight that warnings within the population share the same label,
e.g. if a warning was previously fixed in a file, it is more likely that the other warnings in the same file will be fixed too.

\begin{table}[t]
  \centering
  \caption{The Golden Features studied in prior work~\cite{wanggolden,yangeasy,yangincremental}. A \textit{warning context} is defined~\cite{wanggolden} as the difference of the number of actionable warnings and false alarms divided by the total number of warnings reported in a given method/file, or for a warning pattern. We provide more descriptions of each feature in our replication package~\cite{replication}.}
  \label{tab:golden_features}
  
    \begin{tabular}{|l|l|}
      \hline
 \textbf{Feature type}     & \textbf{Feature} \\
 \hline
 & warning context in method   \\
 & warning context in file    \\
Warning combination & warning context for warning type  \\
 & defect likelihood for warning pattern \\
 & discretization of defect likelihood \\
 & average lifetime for warning type\\
 \hline
 & comment-code ratio \\
 & method depth  \\
 Code characteristics  & file depth \\
 & \# methods in file \\
 & \# classes in package \\
 \hline
 & warning pattern \\
Warning characteristics & warning type  \\
 & warning priority  \\
 & package \\
 \hline
 
 & file age  \\
 File history& file creation \\
 & developers \\
 \hline
 Code analysis & parameter signature \\
 & method visibility \\
 \hline
 Code history & LOC added in file (last 25 revisions) \\
 & LOC added in package (past 3 month) \\
\hline
Warning history & warning lifetime by revision \\
\hline 
\end{tabular}

\end{table}

Further research~\cite{yangeasy} on the Golden Features of Wang et al.~\cite{wanggolden} showed the 
lack of influence of the choice of machine learning model on effectiveness.
They suggested that a linear SVM was optimal since it requires a lower cost of training.
In contrast, while a deep learning approach achieves similar levels of effectiveness, 
it has a longer training time.
Their analysis~\cite{yangeasy} suggested that the detection of false alarms is an intrinsically easy problem.
A different study~\cite{yangincremental} demonstrated that, with the Golden Features, only a small proportion of the dataset has to be labelled to train an effective classifier.
The Golden Features are a subject of our study.
In Section \ref{sec:golden_features}, we analyze them in detail.

\begin{figure}[]

  \includegraphics[width=\columnwidth]{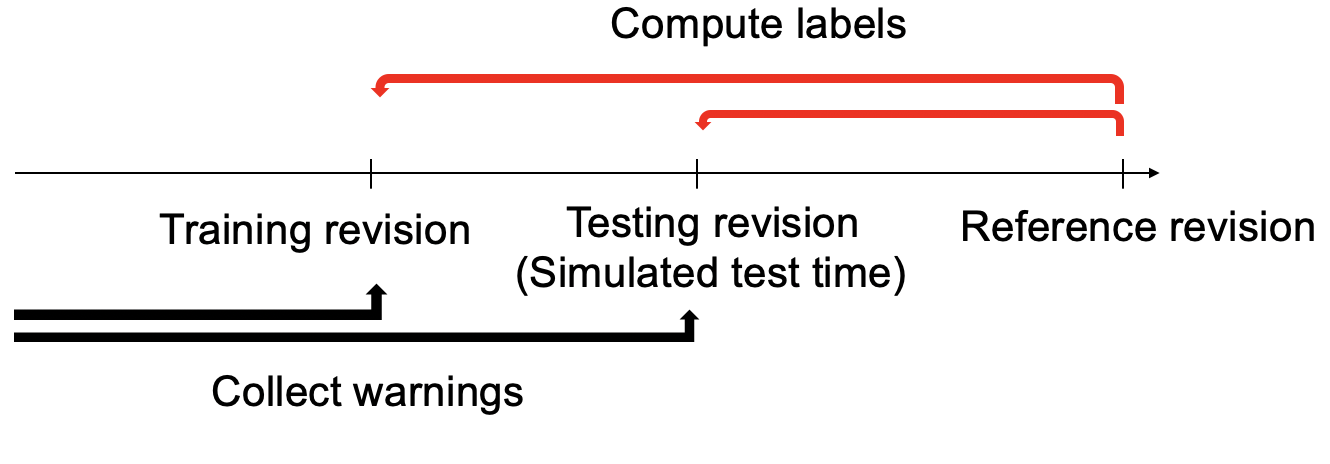}
\centering
\caption{The dataset comprises warnings created before the training and testing revisions. The labels of each warning are determined by the closed-warning heuristic; if a warning is closed at the reference revision and the file has not been deleted, then it is actionable.}
  \label{fig:dataset_construction}
\end{figure}

\textbf{Closed-warning heuristic.} The procedure to construct and label the ground-truth dataset can be visualized in Figure \ref{fig:dataset_construction}.
To assess an approach that detects false alarms,
a  dataset of Findbugs warnings is collected.
While some researchers~\cite{heckman2008establishing,shen2011efindbugs} construct a labelled dataset through manual labelling of the warnings in a single revision, 
other researchers 
collect a dataset through an automatic ground-truth data collection process~\cite{hanam2014finding,heckman2008establishing,wanggolden,yangincremental,yangeasy}.
Data for a \textit{testing revision} and at least one \textit{training revision}, set chronologically before the testing revision, is collected.
This simulates real-world use of the tool, in which training is done on the history of the project, and then used at the time of the testing revision.

Using the \textit{closed-warning heuristic} as the warning oracle, 
each warning in a given revision is compared against a \textit{reference revision} set in the future of the test revision.
Prior studies selected a reference revision set 2 years after the test revision.
If a specific warning is no longer present in the reference revision (i.e., a \textit{closed warning}), the heuristic assumes that the warning is actionable.
If the warning is present in both the given and reference revision (i.e., an \textit{open warning}), then the heuristic assumes that it is a false alarm.
If the file that contains the code with the warning has been deleted, then the warning is labelled \textit{unknown} and is removed from the dataset.
In other words, according to the the closed-warning heuristic, a closed warning is always actionable as long as the file has not been deleted, 
and an open warning is always unactionable.
Other than detecting actionable warnings, researchers have applied the heuristic to identify bug-fixing commits for mining patterns~\cite{liu2018mining,liu2019avatar}.
The heuristic is a subject of our study, and we assess its robustness and its level of agreement with human oracles in Section \ref{sec:dataset_analysis}.

\section{Study Design}
\label{sec:study}
\subsection{Research Questions}

\vspace{0.2cm}\noindent{\bf RQ1. Why do the Golden Features work?} 
This research question seeks to understand the Golden Features.
While previous studies have highlighted their strong results, there has not been an in-depth analysis of their practicality.
We study the Golden Features and the dataset used in the experiments by Wang et al.~\cite{wanggolden} and Yang et al.~\cite{yangeasy}.
We investigate the aspects of the features and dataset 
that allow accurate predictions by the best performing machine learning model, an SVM using the Golden Features.
We replicate the results of the previous studies and validate the predictive power of the Golden Features.
To understand the importance of different features, 
we use LIME~\cite{ribeiro2016should} to narrow our focus down to the features that contribute the most to the predictions.
Afterwards, we switch to increasingly simpler classifiers and analyze the experimental data to better understand why the choice of classifiers did not influence the results in prior studies.

\vspace{0.2cm}\noindent{\bf RQ2. How suitable is the closed-warning heuristic as a warning oracle?}
This research question concerns the suitability of the \textit{closed-warning heuristic} as a warning oracle.
A good oracle should be robust, and its judgments should agree with the analysis of a human annotator.
We investigate the robustness of the heuristic, checking the consistency of labels under different choices of the reference revision.
While previous studies used a 2-years interval between the test revision and reference revision, 
we investigate 
if different conclusions can be reached with a different time interval.
Next, we compute the proportion of closed warnings that human annotators labelled actionable,
and the proportion of open warnings that project developers suppressed as false alarms.

\subsection{Evaluation Setting}

To analyze the performance of machine learning approaches that identify actionable Findbugs warnings,
we use the same metrics as prior studies~\cite{yangincremental,yangeasy,wanggolden}.
A true positive (TP) is an actionable Findbugs warning correctly predicted to be actionable. 
A false positive (FP) is an unactionable Findbugs warning incorrectly predicted to be actionable. 
Note that we use the term \textit{false alarm} to refer to unactionable Findbugs warning. 
A \textit{false positive}, therefore, refers to a false alarm that is incorrectly determined to be an actionable warning.
A false negative (FN) is an actionable warning incorrectly predicted to be a false alarm.
A true negative is an unactionable warning correctly predicted to be a false alarm.

We compute Precision and Recall as follows:
\vspace{0.1cm}
\begin{center}
  $\text {Precision}=\frac{\text { TP } }{\text {TP}+\text {FP}}$ \;\;\;\;
\end{center}
\vspace{0.1cm}
\begin{center}
  $\text {Recall}=\frac{\text { TP } }{\text {TP}+\text {FN}}$
\end{center}
\vspace{0.1cm}

Finally, 
we compute and present F1, the harmonic mean of Precision and Recall.
F1 is known to capture the trade-off between Precision and Recall,
and is used in place of accuracy given an imbalanced dataset.
F1 is computed as follows:

\vspace{0.1cm}
\begin{center}
  $\text {F1}=2 \times \frac{\text { Precision } \times \text { Recall }}{\text {Precision}+\text{Recall}}$
\end{center}
\vspace{0.1cm}


The Area Under the receiver operator
characteristics Curve (AUC) is a measure of the predictive power of a machine learning approach to distinguish between true and false alarms.
Ranging between 0 (worst discrimination) and 1 (perfect discrimination),
AUC is the area under the curve of the true positive rate against the false positive rate, and recommended over accuracy when  the data is imbalanced.
A strawman classifier that always outputs a single label has an AUC of 0.5.

Our dataset consists of projects that were studied by Yang et al.~\cite{yangeasy,yangincremental} and Wang et al.~\cite{wanggolden}.
Similar to previous studies~\cite{yangincremental,yangeasy}, we use one training revision and one testing revision.
We use the same testing revision as previous studies~\cite{yangincremental,yangeasy,wanggolden}.
We train one model for each project.

\section{Analysis of the Golden Features}
\label{sec:golden_features}

To answer the first research question, we investigate the performance of the Golden Features by first using the same dataset used by Yang et al.~\cite{yangeasy}.
The dataset includes two revisions from 9 projects.
The testing revisions are the revisions of the projects on 1 January 2014, 
and the training revision is a revision of the projects up to 6 months before the testing revision.
In total, 31,058 warning instances were obtained by running Findbugs over the training and testing revision.
On average, 14.1\% of the warnings in the dataset were actionable.
Table \ref{table:num_baseline_dataset} shows the breakdown of the warnings.

\begin{table}[t]
  \centering
  \caption{The number of training, testing instances, and the percentage of actionable warnings (Act. \%) in the dataset. The testing revision is the last revision checked into the main branch on 2014-01-01.}
  \label{table:num_baseline_dataset}
  \begin{tabular}{|l|r|r|r|r|r|}
 
 \hline

\multicolumn{1}{|c|}{\textbf{Project}}  & \multicolumn{1}{|c|}{\textbf{Training}}   & \multicolumn{2}{c|}{\textbf{With duplicates}} & \multicolumn{2}{c|}{\textbf{W/o duplicates}}  \\ 
 \cline{3-6} 
 \multicolumn{1}{|c|}{} & \multicolumn{1}{|c|}{}   & \textbf{testing} & \textbf{Act. \%} & \textbf{testing} & \textbf{Act. \%}          \\ 
  \hline 

  ant & 1229 & 1115 &  5\%   & 21 & 71\% \\ 
  cassandra & 2584 & 2601 &  14\%   & 551 &70\% \\ 
  commons & 725 & 786 &  5\%   & 4 & 50\% \\ 
  derby & 2479 & 2507 &  5\%   & 499 & 31\% \\ 
  jmeter & 604 & 613 &  24\%   & 57 & 19\% \\ 
  lucene & 3259 & 3425 &  34\%   & 893 & 59\% \\ 
  maven & 813 & 818 &  3\%   & 149 & 14\% \\ 
  tomcat & 1435 & 1441 &  23\%   & 227 & 41\% \\ 
  phoenix & 2235 & 2389 &  14\%   & 214 & 22\% \\ 
  \hline 
  \end{tabular}
  
\end{table}

\begin{table*}[t]
  \centering
  \caption{Effectiveness of an SVM using the Golden Features after removing the leaked features and removing the duplicate warnings between the training and testing dataset. The numbers in parentheses are the F1 obtained by the baseline classifier that predicts all warnings are actionable.  }
  \label{table:result_baseline_dataset}
  \begin{tabular}{|l|r|r|r|r|r|r|r|r|}
 
 \hline
\multicolumn{1}{|c|}{\textbf{Project}}  & \multicolumn{2}{c|}{\textbf{All Golden Features}} & \multicolumn{2}{c|}{\textbf{\textminus~leaked features}} & \multicolumn{2}{|c|}{\textbf{\textminus~data duplication}} & \multicolumn{2}{c|}{\textbf{\textminus~leak, duplication}} \\ 
 \cline{2-9} 
 \multicolumn{1}{|c|}{}    & \textbf{F1} & \textbf{AUC} & \textbf{F1} & \textbf{AUC} & \textbf{F1} & \textbf{AUC}      & \textbf{F1} & \textbf{AUC}         \\ 
  \hline 

ant  & 0.94 (0.09) & 1.00 & 0.11 (0.09) & 0.67 & - & - & - & - \\
cassandra & 0.92 (0.24) & 1.00 & 0.45 (0.24) & 0.86 &0.9 (0.41) & 0.99 &  0.29 (0.41) & 0.54 \\
commons & 0.65 (0.10) & 0.99 & 0.16 (0.10) & 0.65 & 0.75 (0.25) & 0.97 & 0.11 (0.25) & 0.49 \\
derby & 0.95 (0.09) & 1.00 & 0.39 (0.09) & 0.93 & 0.97 (0.28) & 0.97 & 0.30  (0.28) & 0.59 \\
jmeter & 0.94 (0.38) & 0.99 & 0.53 (0.38)  & 0.76 & 1.00 (0.14) & 1.00 &  0.25 (0.14) & 1.00 \\
lucene-solr & 0.87 (0.51) & 0.97 & 0.59 (0.51) & 0.74 & 0.87 (0.53)  & 0.98  & 0.23 (0.53)& 0.62 \\
maven & 0.86  (0.07) & 1.00 & 0.27  (0.07)  & 0.9 & 0.95 (0.24) & 0.99 & 0.27 (0.24) & 0.58 \\
tomcat & 0.93 (0.37) & 1.00 & 0.48 (0.37)  & 0.73 & 0.95 (0.70) & 1.00 & 0.65 (0.70) & 0.39 \\
phoenix & 0.89 (0.25) & 1.00 & 0.42 (0.25) & 0.78 & 0.83 (0.37) & 0.99 & 0.40 (0.37) & 0.63 \\
\hline 
Average & 0.88 (0.23) & 1.00 & 0.38 (0.23) & 0.76 & 0.90 (0.37) & 0.99 & 0.31 (0.37) & 0.59 \\

  \hline 
  \end{tabular}
  
\end{table*}

We successfully replicate the performance observed in the experiments of Yang et al.~\cite{yangeasy} and Wang et al.~\cite{wanggolden}, 
obtaining high AUC values of up to 0.99.
An average F1 of 0.88 was obtained, with F1 ranging from 0.65 to 0.95.
Table \ref{table:result_baseline_dataset} shows our experimental results.

Yang et al.~\cite{yangeasy} found that the dataset was intrinsically easy as the data was inherently low dimensional.
To further analyze their findings, we used tools from the field of explainable AI, in particular LIME~\cite{ribeiro2016should},
to identify the most important features contributing to each prediction.
LIME is an explanation technique that identifies the most important features that contributed to an individual prediction.
To identify the most important features, 
we sampled 50 predictions made by the Golden Features SVM, 
and used LIME to identify the top features contributing to the predictions.
We found that two features, \textbf{warning context of file} and \textbf{warning context of package}, 
appeared in the top-3 features of every prediction.

\textbf{Warning context and defect likelihood features.} 
On analyzing
the source code of the feature extractor developed by Wang et al.~\cite{wanggolden},
we found a subtle data leak in the implementation of the warning context and defect likelihood features.
These features utilize findings from previous studies~\cite{kremenek2004correlation}
that found that the  warnings within a population (e.g. warnings in the same file) tend to be homogenous; 
if one warning is a false alarm, then the other warnings in the same population tend to be false alarms as well.
Including \textbf{warning context of file} and \textbf{warning context of package}, there 
are another 3 features computed similarly (\textbf{warning context of warning type}, \textbf{defect likelihood for warning pattern}, \textbf{Discretization of defect likelihood for warning pattern}).
At a  high level, the warning context features are computed as follows: \\
\begin{center}
  $ \frac{| W^{\mathit{actionable}}_{\mathit{relevant}} | - | W^{\mathit{false~alarm}}_{\mathit{relevant}} | }{| W_{\mathit{relevant}} |} $
  \end{center}
  \vspace{0.1cm}

$W_{\mathit{relevant}}$ refers to the set of warnings  relevant to the feature type.
For example, $W_{\mathit{relevant}}$ of \textbf{warning context of file} considers the warnings that are reported in a given file,
while $W_{\mathit{relevant}}$ of \textbf{warning context of warning type} considers all warnings for the given category of patterns (e.g. \texttt{STYLE}, \texttt{INTERNATIONALIZATION}).
Note that a warning pattern refers to a specific bug pattern in Findbugs (e.g. ``ES\_COMPARING\_STRINGS\_WITH\_EQ''), and a warning type is a category of patterns.
The \textbf{defect likelihood for warning pattern}~\cite{shen2011efindbugs} feature computes the proportion of warnings that were actionable out of all warnings with the given bug pattern, $p$:
\begin{center}
  $ D(p) = \frac{| W^{\mathit{actionable}}_{\mathit{relevant}} |}{| W_{\mathit{relevant}} |} $
  \end{center}
  \vspace{0.1cm}

The \textbf{discretization of defect likelihood for warning type}~\cite{shen2011efindbugs} feature, computed for each type/category $T$ of bug patterns,
is a measure of the difference in defect likelihood from the defect likelihood of $T$ for each bug pattern in the category: 

\begin{center}
  $ \frac{1}{|T| - 1} \sum_{p \in T} (D(p) - D(T))^{2} $
  \end{center}
  \vspace{0.1cm}

\begin{figure}[]

  \includegraphics[width=\columnwidth]{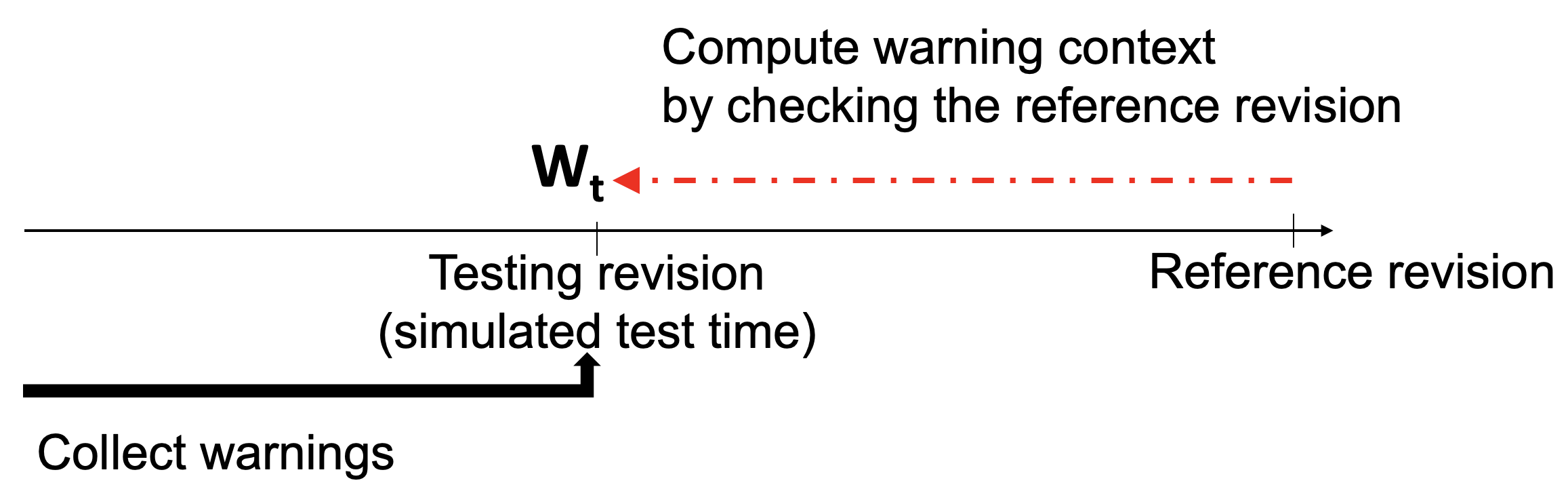}
\centering
  \caption{The warning context and defect likelihood features use labels derived through the closed-warning heuristic, using information from the reference revision, chronologically in the future of the test revision. In a realistic setting, this information will not be present at test time.}
  \label{fig:data_leak}
\end{figure}

The five warning context and defect likelihood features require information about the actionability of each warning in the population of warnings considered.
A data leakage occurs when the classifier utilizes information that is unavailable at the time of its predictions~\cite{tu2018careful,kaufman2012leakage}.
As shown in Figure \ref{fig:data_leak}, 
while the ratio of actionable warnings are computed over the warnings reported in the past (the black line in Figure \ref{fig:data_leak}),
the closed-warning heuristic to determine the ground-truth label of a warning (the red lines in Figure \ref{fig:dataset_construction} and Figure \ref{fig:data_leak}) is utilized to determine if these warnings were actionable.
To compute the warning context of a given warning, $W_{t}$ in the testing revision,
the labels of all warnings in the population of warnings (e.g. all warnings in the same file), 
including $W_{t}$, are obtained based on comparison to the reference revision.

Since the ground-truth label is also obtained based on comparison to the reference revision,
the ground-truth label is \textit{inadvertently leaked} into the computation of the warning context.
This is not a realistic assumption in practice; at test time, the ground-truth label of the warning context of $W_{t}$ is the target of the prediction.
While checking if a warning will be closed 2 years in the future is possible within an experiment,
there is no way to check if the warnings will be closed 2 years into the future in practice.
Table \ref{table:result_baseline_dataset} shows the large drop in F1, from an average of 0.88 to 0.38, when these five features are dropped.
We refer to these features as \textit{leaked features}.

\textbf{Baseline using data leakage.} Data leakage leads to an experimental setting that overestimates the effectiveness of the classifier under study~\cite{tu2018careful,kaufman2012leakage}.
In Table \ref{table:result_baseline_dataset_different_models}, 
we show that a baseline equivalent to the Golden Features can be developed using only the five leaked features.
Using just the leaked features with an SVM, we construct a baseline that achieves performance comparable to the use of the Golden Features.
An SVM using the leaked features has a Precision of 0.79, about 0.10 lower than the Golden Features SVM,
however, they achieve identical Recall of 0.94, which results in an F1 of 0.83, just 0.05 lower than the Golden Features.
This indicates that the strong performance of the Golden Features in the experiments 
depends largely on the leaked features,
and is an optimistic estimate of their  effectiveness.
 
\vspace{0.1cm}\noindent\fbox{%
\parbox{\columnwidth}{%
The computation of the \textit{warning context} and \textit{defect likelihood} features caused data leakage, as it used labels determined by comparison against the reference revision, chronologically in the future of the testing time.
}%
}
\vspace{0.1cm}

\textbf{Data duplication.} Next, 
we progressively selected simpler machine learning models and surprisingly,
found that a k-Nearest Neighbors (kNN) classifier performs effectively.
In particular, we found a surprising trend where the lower values of $k$ led to better results.
The results of the experiment where we iteratively lowered $k$ to consider in the prediction are shown in Table  \ref{table:result_baseline_dataset_different_models}.

Surprisingly, a kNN classifier with k=1 (i.e., only one neighbor is considered to make a prediction) produces the best result,
obtained a Precision of 0.87, a Recall of 0.90, with an F1 of 0.84.
With k=1, the classifier was selecting a single most similar warning in the training dataset.
In typical usage of kNN, a low value of $k$ may cause the classifier to be influenced by noise and outliers,
which makes the strong results surprising.
To analyze the results further, we observed that the number of training (15,363) and testing instances (15,695) were similar,
and we investigated the data carefully.
We found that many testing instances appeared in both the training and testing dataset.

The data duplication was caused by the data collection process, 
in which all warnings produced by Findbugs for both the training and test revisions were included in the training and testing dataset.
Say we have a warning at the training revision, determined to be open and, therefore, unactionable by the closed-warning heuristic.
In other words, the warning remained open in the period before the training revision to the reference revision.
Then, the warning would certainly be opened at the testing revision, which is chronologically before the reference revision but after the training revision.
Likewise, if we have a warning 
only closed after the testing revision,
but was open during the testing revision,
then the same warning would be present at both the training and testing revision with the same ``actionable'' label.
Consequently, a large number of warnings appear in both the training and testing dataset.
This contributes to an unrealistic experimental setting.

\begin{table}[t]
  \centering
  \caption{Average Precision (Prec.), Recall, and F1 of various approaches on the original dataset by Yang et al.}
  \label{table:result_baseline_dataset_different_models}
  \begin{tabular}{|l|r|r|r|}
 
 \hline

\textbf{Technique}           & \textbf{Prec.} & \textbf{Recall} & \textbf{F1} \\ 
  \hline 
    Golden Features SVM               & 0.84 & 0.94 & 0.88              \\ 
    \textminus~leaked features       & 0.26 & 0.70 & 0.38              \\ 
    \textminus~data duplication      & 0.88 &  0.93 & 0.90              \\ 
    \textminus~data duplication and leaked features      & 0.27 &  0.57   & 0.31              \\ 
    +~reimplemented leaked features  & 0.32 & 0.57 & 0.38    \\ 
    \hline 
    Golden Features kNN (with k=10)  & 0.91 & 0.57 & 0.68 \\
    Golden Features kNN (with k=5)  & 0.86 & 0.72 & 0.78 \\
    Golden Features kNN (with k=3)  & 0.87 & 0.78 & 0.82 \\
    Golden Features kNN (with k=1)  & 0.87 & 0.90 & 0.84 \\
    Only leaked features SVM            &0.79 & 0.94 & 0.83         \\ 
    Repeat label from training dataset  & 0.72& 0.80 & 0.75   \\     
\hline 
  \end{tabular}
  
\end{table}

\textbf{Baseline using duplicated data.} Data duplication creates an artificial experimental setting that inflates performance metrics~\cite{allamanis2019adverse}.
To confirm that the data duplication contributes to the ease of the task,
we construct a weak baseline, a dummy classifier, that leverages the duplication of testing data in the training  dataset.
Given a warning from the testing dataset, 
the classifier
heuristically identifies the same warning from the training dataset by searching for a training warning based on the class name (e.g. ``BooleanUtils'')
and bug pattern name (e.g. ``ES\_COMPARING\_STRINGS\_WITH\_EQ'').
If there are multiple warnings with the same class name and bug pattern name, a random training instance is selected from among them.
The classifier then outputs the label of the training instance.
If there is no training instance with the same class and bug pattern type, 
then the classifier defaults to predicting that the warning is a false alarm, which is the majority class label.

Table \ref{table:result_baseline_dataset_different_models} shows the comparison of various approaches, including the baseline approaches, 
on the dataset.
The dummy classifier achieves strong performance, achieving a Precision of 0.72, a Recall of 0.80,
and an F1 of 0.75. 
While the dummy classifier underperforms the model using the leaked features, it outperforms the Golden Features SVM without the leaked features.
This indicates that using just two attributes, (1) the class name and (2) the bug pattern of the warning, is enough to obtain strong performance on a dataset with data duplication.
Therefore, we conclude that the data duplication between the training and testing dataset contributes to the strong performance observed in previous studies.

\begin{figure}[]

  \includegraphics[width=\columnwidth]{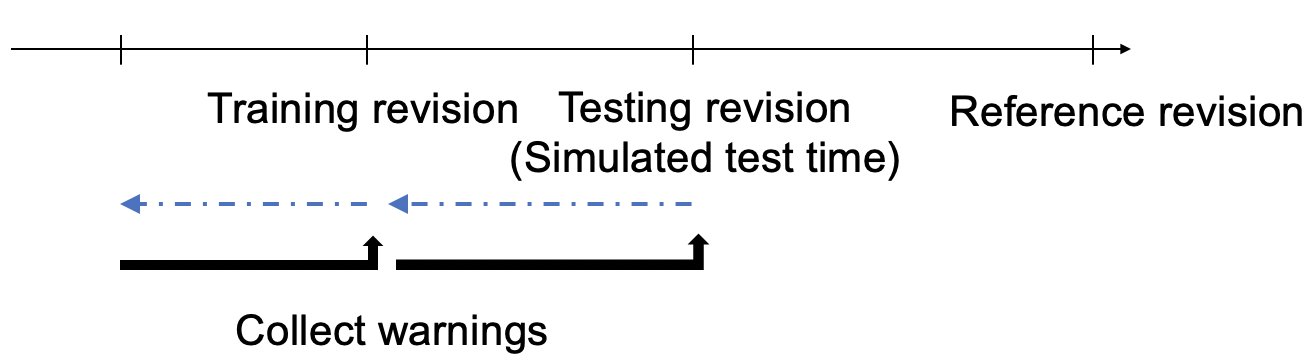}
\centering
  \caption{We reimplemented the leaked features. The reimplemented features use only information (represented by the blue, dashed lines) available at the present (i.e., either the training or test revision) to determine if a warning (i.e., created before the training or test revision) has been closed. Under this setting, no information from the reference revision is used for making predictions.}
  \label{fig:reimplemented_features}
\end{figure}

The experimental results are summarized in Table \ref{table:result_baseline_dataset_different_models}.
With both the leaked features and duplicated data, the average F1 was 0.88.
After the data leakage features are removed, F1 decreased to 0.38.
After removing the duplicated data, F1 decreases further to 0.31.
The average project's AUC decreased from 1.00 to 0.59. 
In comparison, using a strawman baseline that predicts that every warning is actionable produces an F1 of 0.52 (with an AUC of 0.5).

\vspace{0.1cm}\noindent\fbox{%
\parbox{\columnwidth}{%
All warnings reported by FindBugs on both the training and testing revisions were included in the datasets. Warnings reported at the training revision may still be reported at the testing revision, leading to data duplication between the training and testing dataset.
}%
}
\vspace{0.2cm}

\textbf{Experiments under a more realistic setting.} 
To better understand the performance of the Golden Features SVM, 
we ran another experiment where the two issues of data leakage and data duplication have been fixed.
First, we deduplicated the test data from the training dataset.
Instead of including all warnings in the testing revision, we only consider new warnings introduced between the time 
after the training revision and before the testing revision.
Figure \ref{fig:reimplemented_features} shows our procedure. 
As compared to the previous dataset construction process in Figure \ref{fig:dataset_construction}, 
only the warnings created after the training revision and before the testing revision are used for testing.
This better reflects real-world conditions 
where all warnings prior to usage are used for training, but none of the testing data involves warnings that have already been classified.
In total, the number of warnings in the testing revisions decreased from a total of 15,695 to 2,615 after deduplication.
Without the duplicated data and without using the leaked features, 
the average F1 drops from 0.88 to 0.31 as seen in Table \ref{table:result_baseline_dataset}.

Next, we reimplemented the leaked features to investigate the effectiveness of Golden Features SVM.
To prevent data leakage, we modified the definition of the leaked features.
Figure \ref{fig:reimplemented_features} visualizes the computation of the warning context and defect likelihood features.
Instead of considering all warnings, 
we consider only warnings that were introduced in the 1 year duration before the training or testing revision. 
Instead of using the reference revision, we use the given revision (i.e., either the training or testing revision) to determine if the warning was closed.
A warning is closed at a given revision if Findbugs does not report it.
In other words, for the training revision, only the warnings created within the past year before the training revision are considered.
For testing, only the warnings created within one year before the testing revision are considered. 
A time interval of 1 year was selected in contrast to the study by Wang et al.~\cite{wanggolden}, which used time intervals of up to 6 months.
Unlike Wang et al.~\cite{wanggolden}, for the testing revision, we consider only warnings created after the training revision to prevent data duplication.
Consequently, we found fewer newly created warnings in the short time interval between the training and testing revisions.

Note that after reimplementing the warning context and defect likelihood features, we could not run the experiments for the project Phoenix
as we faced many difficulties building old versions of the project.
Moreover, their revision history did not go back beyond 3 years, 
required for computing the warning context and defect likelihood features for the training revision.
This limitation is not present for Wang et al.~\cite{wanggolden}, as they compute the features by checking if the given warning is closed in the reference revision, set in the future of the test revision (causing data leakage).
As such, we omit Phoenix for the rest of the experiments.

Table \ref{table:result_baseline_dataset_different_models} shows the performance of the Golden Features SVM using the reimplemented features.
Without the leaked features, the Golden Features SVM achieves an F1 of 0.38.
Even with the reimplementation of the leaked features, the Golden Features SVM underperforms the strawman baseline, which predicts all warnings are actionable,
with an F1 of 0.43.
However, it has an AUC of 0.59, greater than 0.5, indicating that the Golden Features are better than random and have some predictive power.

\vspace{0.2cm}\noindent\fbox{%
\parbox{\columnwidth}{%
\textbf{Answer to RQ1:} After removing the data leakage and data duplication, our experimental results indicate that the Golden Features SVM underperforms the strawman baseline, although its AUC (> 0.5) suggests that the Golden Features have some predictive power.
}%
}
\vspace{0.1cm}

\section{Analysis of the Closed-Warning Heuristic }
\label{sec:dataset_analysis}

Next, 
given that the quality and realism of the dataset heavily influences the evaluation of the Golden Features SVM,
we perform a deeper analysis of the construction of the ground-truth dataset.
In previous studies~\cite{wanggolden,yangeasy,yangincremental}, the warning oracle 
is the \textit{closed-warning heuristic};
a warning is heuristically determined to be actionable if it was closed (i.e., reported by Findbugs in a revision but was not reported by Findbugs in the reference revision, and the file was not deleted),
and is a false alarm if it was open (i.e. reported by Findbugs on both the training/test and reference revision).

In the first part of our analysis, we investigate the consistency in the warning oracle given a change in the reference revision.
Next, we check if the warning oracle produces labels that human users would agree with.
To do so, we first determine if human annotators consider closed warnings as actionable warnings.
In addition, we match open warnings against Findbugs filter files in projects where developers have configured the filters for suppressing false alarms.
Finally, we observe if cleaner data increases the effectiveness of the Golden Features SVM.

\subsection{Choosing a different reference revision}
\label{sec:different_ref}
We perform a series of experiments to determine how the time interval between the test revision and the selected reference revision influences the ground-truth label of the warnings.
We hypothesize that the longer the time interval between the test and reference revision, the greater the proportion of closed warnings.
Based on the closed-warning heuristic, this would cause more warnings to be labelled actionable.
If so, the lack of consistency in labels should call the robustness of the heuristic into question.
If many bugs are fixed only after many years, then an open warning at any given time may, in fact, be actionable.
Besides that, if changing the reference revision leads us to a different conclusion about the Golden Features SVM,
then it limits the level of confidence that researchers can have in the experimental results.

In our experiments, we use three reference revisions set two, three, and four years after the test revision.
By switching the reference revision, we observe changes in the average actionability ratio.
While the actionability ratio remained consistent for the 4 out of 8 projects,
the actionability ratio increased by over 10\% for the other 4 projects, as seen in Table \ref{table:num_2016_2017_2018}.
Overall, the average actionability ratio increased by 14\% when varying the time interval between the test and reference revision from 2 to 4 years.
Considering all projects, we performed a Wilcoxon signed-rank test and found that the change in actionability ratio is statistically significant (p-value=0.03 < 0.05).

In terms of the effectiveness of the Golden Features SVM, its average F1 increased from 0.39 to 0.57, as seen in Table \ref{table:num_2016_2017_2018}.
Considering all projects, the Golden Features SVM underperformed the strawman baseline.
Our experiments showed some variation of the Golden Features SVM's effectiveness given a change in the reference revision.
For instance, the Golden Features SVM achieved a low F1 of 0.06 in Derby when the time interval between the test and reference revision was 2 years,
but had a high F1 of 0.72 with a time interval of 4 years.

By changing reference revisions, the problem exhibits different characteristics. 
Using a reference revision 4 years after the test revision, 
actionable warnings would be the majority class, while they were the minority class when using the other reference revisions.
4 of 8 projects have an AUC that flipped from one side of 0.5 to the other (e.g. the Golden Features SVM's AUC is under 0.5 on Derby given a 2-years interval, but the AUC increases above 0.5 given a 4-years interval).
In short, different conclusions about the task and the effectiveness of the Golden Features may be reached.

\begin{table*}[t]
  \centering
  \caption{The number of training, testing instances, and the percentage of actionable warnings (Act. \%) in the dataset when varying the reference revision. The numbers in parentheses are the F1 obtained by the baseline classifier that predicts all warnings are actionable. The testing revision is the last revision checked in to the main branch before 2014-01-01. }
  \label{table:num_2016_2017_2018}
  \begin{tabular}{|l|r|r|r|r|r|r|r|r|r|r|}
 
 \hline
 \multicolumn{1}{|c|}{\textbf{Project}}  & \multicolumn{1}{|c|}{\textbf{\# testing}}  & \multicolumn{3}{c|}{\textbf{2 years}} & \multicolumn{3}{c|}{\textbf{3 years}} & \multicolumn{3}{c|}{\textbf{4 years}} \\ 
 \cline{3-11} 
 \multicolumn{1}{|c|}{} & \textbf{instances}     & \textbf{Act. \%}  &\textbf{F1} & \textbf{AUC} &   \textbf{Act. \% } & \textbf{F1} & \textbf{AUC} & \textbf{Act. \% } & \textbf{F1} & \textbf{AUC}  \\ 
  \hline 
 ant   & 21 & 24 & 0 (0.38)	& 0.43 &    43  &  0.13 (0.60) & 0.48  & 43 & 0 (0.60) & 0.32    \\ 
 cassandra & 551&  41 & 0.56 (0.59)	& 0.52  &   46  & 0.61 (0.63) & 0.41  & 43   & 0.58 (0.60) & 0.48  \\ 
 commons &  4 & 50  &  0.66 (1.00) & 	1.00 &  50  & 1.00 (0.67) & 1.00 &  50  &1.00 (0.67) & 1.00   \\ 
 derby &  489 & 10   & 0.06 (0.18)	& 0.33 &  59 & 0.58 (0.74) & 0.46  & 66   & 0.72 (0.80) & 0.52  \\ 
 jmeter & 57 & 17  & 0.13 (0.16)	& 0.58 &    26 &  0.12 (0.27) & 0.4  & 91   & 0.71 (0.95) & 0.52  \\ 
 lucene & 993 &  44  & 0.56 (0.62)	& 0.58  &   49  & 0.58 (0.66) & 0.57  & 67 & 0.63 (0.8) & 0.53    \\ 
 maven & 149 & 17  &   0.25 (0.29)	& 0.41  & 16  & 0.27 (0.28) & 0.44  & 16  & 0.27 (0.28) & 0.44  \\ 
 tomcat & 226&  42  & 0.53 (0.59)	& 0.51 &  61  & 0.51 (0.57) & 0.48 & 52  & 0.64 (0.69) & 0.54   \\ 
 
\hline 
 Average & 311 & 40 & 0.39 (0.43)	& 0.54 & 42 & 0.48 (0.55)  & 0.53  & 54 & 0.57 (0.67)  & 0.54 \\
\hline 

\end{tabular}

\end{table*}

  \vspace{0.1cm}\noindent\fbox{%
    \parbox{\columnwidth}{%
    Changing the reference revision may affect the distribution of the actionable warnings,
    which may impact the conclusions reached from experiments on the effectiveness of the Golden Features SVM.
    }%
}
\vspace{0.1cm}

\subsection{Unconfirmed actionable warnings}
\label{sec:unconfirmed_positives}
Next, we investigate if closed warnings are truly actionable warnings.
A warning could be closed due to several reasons.
Code containing the warning could be deleted or modified while implementing a new feature,
and the warning may only be closed incidentally.

To further understand the characteristics of closed warnings, 
and to determine how likely is a closed warning an actionable warning,
we sampled 1,357 warnings (which is more than the statistically representative sample size of 384 warnings) that were closed.
Two authors of this study independently analyzed each warning to determine if 
they were removed for a bug fix.
If the warning was closed due to code changes unrelated to the warning, then we do not consider the warning as actionable.
If the code containing the warning was modified such that it was not easily discernible if the warning was closed with the intention of fixing the warning, then we consider it ``unknown''.
If the original version of the code had any comments indicating that Findbugs reported a false alarm 
(e.g. explaining the reason that a seemingly buggy behavior was expected behavior), 
then we consider the warning a false alarm.
When the labels differed between the annotators, they discussed the disagreements to reach a consensus.
We computed Cohen's Kappa to measure the inter-annotator agreement 
and obtained a value of 0.83, which is considered as strong agreement~\cite{landis1977measurement}.

\begin{figure}[]

  \includegraphics[width=\columnwidth]{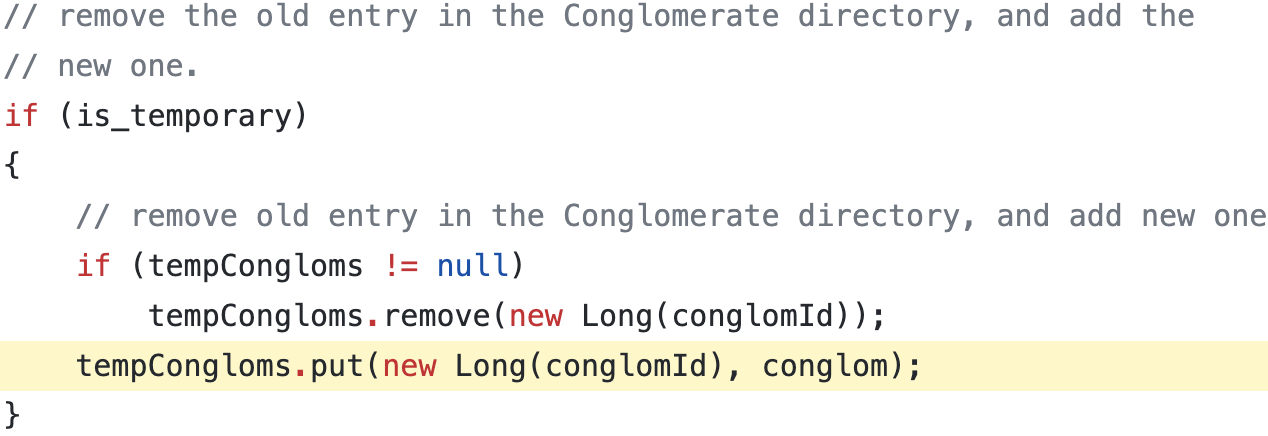}
\centering
\caption{Example of code that Findbugs reports a warning on. Findbugs warns against using \texttt{new Long}, recommending the more efficient \texttt{Long.valueOf} to instantiate a \texttt{Long} object.}
  \label{fig:closed_warning}
\end{figure}

Finally, after labelling, 176 (13\%) of the heuristically-closed warnings were considered as false alarms.
Another 520 warnings (38\%) were categorized as ``unknown''.
Lastly, 660 (49\%) warnings were still considered actionable after labelling.

\begin{figure}[]

  \includegraphics[width=\columnwidth]{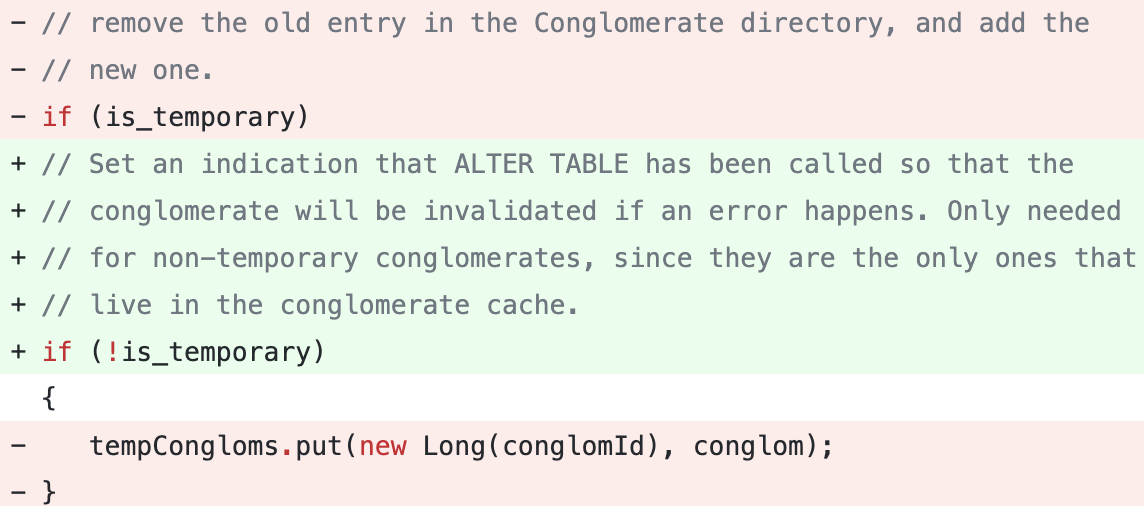}
\centering
\caption{The warning from Figure \ref{fig:closed_warning} is removed through a change in functionality, unrelated to the warning otherwise.}
  \label{fig:unknown_fix}
\end{figure}

For an example of a warning labelled ``unknown'', 
Figure \ref{fig:closed_warning} shows a fragment of code where
Findbugs complains about the use of the \texttt{Long} constructor, 
indicating that \texttt{Long.valueOf} would be more efficient.
Even though the warning is removed in the reference revision, 
the entire functionality of the code fragment was changed as shown in Figure \ref{fig:unknown_fix}.
In such cases, we label the warning as ``unknown'' instead of ``actionable'' or a ``false alarm'',
as there is no evidence that the warning was fixed or ignored.
We consider that the warning was removed incidentally, and that the annotators are unable to accurately label the warning.

While the closed-warning heuristic considered that a warning could be removed through the deletion of a file,
it does not consider other cases where a warning could be incidentally removed through code modification that does not fix the bug indicated by the warning,
Our results indicate 
that more information should be considered, 
and 
that the heuristic may not be sufficiently robust.

\vspace{0.1cm}\noindent\fbox{%
    \parbox{\columnwidth}{%
    Only 47\% of closed warnings were labelled actionable by human annotators,
    implying that many closed warnings are not actionable.
    Many closed warnings were only closed incidentally.
    }%
}
\vspace{0.1cm}

\subsection{Unconfirmed false alarms}
\label{sec:unconfirmed_negatives}
Our findings from Section \ref{sec:different_ref} indicate the possibility that 
some actionable warnings would only be closed given a longer time interval between the test revision and the reference revision.
This may reflect real-world conditions, where developers may not prioritize reports from ASATs
and may take a long time before inspecting them.
Thus, open warnings may be actionable warnings that the developers would fix with enough time.
We run an experiment to understand this effect, focusing on projects
that have shown evidence of using Findbugs.
In this experimental setup, we remove open warnings that are not confirmed by the project developers to be false alarms.

Some projects, which use Findbugs in their development process, configure a Findbugs filter file~\cite{findbugs_filter} 
for indicating false alarms.
The filter file allows developers to suppress warnings of specific bug patterns on the indicated files.
Developers may add warnings to the Findbugs filter file after inspecting the warnings and identifying false alarms.
On projects that have created and maintained a Findbugs filter file, 
we assume that a developer would either fix the buggy code or update the filter file after inspecting a warning.
If so, then an open warning that is not matched by the Findbugs filter file may not be a false alarm, 
but has not been inspected by a developer.
These open warnings could be false alarms, but they may also be warnings that developers would act on after inspecting them.
If an open warning matches the filter, then it has been confirmed by the developers to be a false alarm.

To investigate the proportion of open warnings that are confirmed to be false alarms by project developers, 
we identified 3 projects (JMeter, Tomcat, Commons-Lang) that have already configured the Findbugs filter file from Wang et al.'s dataset~\cite{wanggolden}, 
used in the preceding experiments.
Next, we searched GitHub for mature projects that showed evidence of using Findbugs and have configured a Findbugs filter file.
Using the GitHub Search API, we looked for XML files containing the term \texttt{FindbugsFilter}, which is a keyword used in Findbugs filter files,
in projects that were not forks,
filtering out projects with less than 100 stars or had less than 10 lines in the Findbugs filter file.
We obtained 8 projects.

\begin{table}[t]
  \centering
  \caption{Number of open warnings in each project matched by their Findbugs filter file. If a warning was filtered, it indicates that the project's developers consider it a false alarm.}
  \label{table:num_dev_confirmed}
  \begin{tabular}{|l|r|r|r|}
 
 \hline
                 
 \textbf{Project}           & \textbf{\# open warnings} & \textbf{\# filtered} & \textbf{\% filtered}  \\
  \hline 
  jmeter  & 710 & 6 & 1\% \\
  tomcat  & 1624& 9 & 1\% \\
  commons-lang &106 & 19 & 18\% \\
  flink  & 4934&  4754 & 96\% \\
  hadoop  & 3053& 269 & 9\% \\
  jenkins  & 1212 & 178 & 15\% \\
  kudu  & 1873 & 464 & 25\% \\
  kafka  & 4668 & 2993 & 64\% \\
  morphia  & 65& 0 & 0\% \\
  undertow  & 347 & 113 & 33\%  \\
  xmlgraphics-fop  &949  & 909 & 96\% \\
  \hline 
  Average (Mean) & 1666 & 818 & 31\% \\
  Average (Median) & 1212 & 178 & 18\%  \\
  \hline
  \end{tabular}
  
\end{table}

The statistics of the warnings reported by Findbugs on the projects are displayed in Table \ref{table:num_dev_confirmed}.
On average, 31\%  of the open warnings (a median of 18\%) are matched by the Findbugs filter configured by the developers, 
although the proportion varies for each project.
Our results suggest that the majority of open warnings remain uninspected by developers.

\vspace{0.1cm}
\noindent\fbox{%
    \parbox{\columnwidth}{%
       On average, only 31\% of open warnings have been explicitly indicated by developers to be false alarms,
       suggesting that only a minority of open warnings are false alarms. 
       While the rest of the open warnings could be false alarms, they could also be actionable warnings that have not been inspected yet. 
    }%
}
\vspace{0.1cm}

\begin{table}[t]
  \centering
  \caption{Effectiveness of the Golden Features SVM after removing unconfirmed actionable warnings and false alarms. Act. \% refers to the proportion of actionable warnings. The numbers in parentheses are the F1 of the dummy baseline, which predicts that all warnings are actionable.}
  \label{table:result_clean_data}
  \begin{tabular}{|l|r|r|r|}
 
 \hline
                 
 \textbf{Dataset}           &  \textbf{Act. \%} & \textbf{F1} & \textbf{AUC}          \\
  \hline 
     Original dataset~\cite{yangeasy,yangincremental} & 39.9 & 0.39 (0.43) & 0.54 \\
     \textminus~unconfirmed actionable warnings     & 40.0  &  0.61 (0.57) & 0.66  \\ 
     \hline
     Projects using Findbugs  & 38.0  & 0.43 (0.44) & 0.62 \\
     \textminus~unconfirmed false alarms &  40.0 & 0.41 (0.46) & 0.60 \\
  \hline
  \end{tabular}
  
\end{table}

Next, we investigate the impact of the unconfirmed actionable warnings and false alarms on the Golden Features SVM.
We hypothesize that cleaning up the data will improve its effectiveness.

To study the impact of unconfirmed actionable warnings,
we used the dataset of warnings from the projects by Wang et al.~\cite{wanggolden} and Yang et al.~\cite{yangeasy,yangincremental}.
These projects were the same projects studied earlier in Section \ref{sec:unconfirmed_positives}.
We construct a dataset of warnings with only the warnings confirmed by the human annotators to be actionable warnings.
We randomly sampled a subset of open warnings to retain a similar actionability ratio.

For evaluating the effect of unconfirmed false alarms,
we used the warnings from the projects that used Findbugs (from Section \ref{sec:unconfirmed_negatives}).
However, we omit 4 projects (JMeter, Tomcat, Hadoop, Morphia) where less than 10\% of open warnings matched the filter file,
as the low percentage may indicate that the Findbugs filter files are not kept up to date in these projects.
From the other projects, only open warnings that match the filter file are included. 
We sampled a subset of closed warnings to retain a similar actionability ratio.

The outcome of our experiment is shown in Table \ref{table:result_clean_data}.
Removing unconfirmed actionable warnings led to an increased AUC from 0.54 to 0.68, and an increased F1 from 0.39 to 0.64.
This outperforms the strawman baseline which has an F1 of 0.57,
suggesting that cleaner data may increase the effectiveness of the Golden Features SVM.
However, removing unconfirmed false alarms did not help.
The results may indicate that cleaner data may help and
removing unconfirmed actionable warnings, which is the minority class, may have a positive effect on the effectiveness of a classifier.

\vspace{0.2cm}\noindent\fbox{%
    \parbox{\columnwidth}{%
       \textbf{Answer to RQ2:} The closed-warning heuristic may not be an appropriate warning oracle.
       It lacks consistency with respect to the choice of reference revision, which may affect the findings reached from the experimental results.
       Moreover, the heuristic conflates closed warnings for actionable warnings and open warnings for false alarms.
       We find having cleaner data by removing unconfirmed actionable warnings can boost the performance of the Golden Features SVM.
    }%
}
\vspace{0.1cm}

\section{Discussion}
\label{ref:discussion}

\subsection{Lessons Learned}

\textbf{To detect actionable warnings, the Golden Features alone are not a silver bullet.}
Our results indicate that the performance of the Golden Features SVM is not almost perfect, with only marginal improvements over a strawman baseline that always
predicts that a warning is actionable.
Our study motivates the need for more work. 
Future work should  explore more features and techniques, 
including pre-processing methods (e.g. SMOTE) and other machine learning methods (e.g. semi-supervised learning).

\textbf{All that glitters is not gold; it is essential to qualitatively analyze and understand the reasons for seemingly strong performance.}
Despite achieving excellent performance, the Golden Features have subtle bugs related to data leakage and data duplication.
This emphasizes the importance of a deeper analysis of experimental results, and both quantitative and qualitative analysis are essential.
We call for the need for more replication studies, as such works can highlight opportunities and challenges for future work. 
Our work reemphasizes the need to compare both existing and newly proposed techniques to simple baselines~\cite{fu2017easy}.

\textbf{The closed-warning heuristic for generating labels allows a large dataset to be built  but is not enough for building a benchmark.}
Our work sheds light on the limitations of the closed-warning heuristic,
suggesting that it may not be sufficiently accurate;
warnings may be closed incidentally, and
actionable warnings may stay open for years before they are closed.

As a benchmark is essential for charting research direction~\cite{sim2003using}, 
the construction of a representative dataset is important.
Several studies have proposed similar processes relying on the closed-warning heuristic to build a ground-truth dataset~\cite{wanggolden,hanam2014finding,kim2007prioritizing,heckman2013comparative}, while others have relied on manual labelling~\cite{heckman2008establishing,heckman2009model,shen2011efindbugs,ruthruff2008predicting,yuksel2013automated,koc2019empirical}.
Heuristics enables automation, allowing for a  dataset of a greater scale.
However, heuristics may not be robust enough.
On the other hand, solely labelling warnings through manual analysis is not scalable and may be subject to an annotator's bias.
We suggest that datasets proposed in the future should rely on \textbf{both} heuristics and manual labelling;
apart from its greater scale, the closed-warning heuristic enables rich information to be gathered from the activities of the developers to help the manual labelling process.
For example, code commits provide richer information, such as the commit message, simplifying the task for human annotators.
In contrast, prior studies~\cite{heckman2008establishing,heckman2009model,shen2011efindbugs,ruthruff2008predicting,yuksel2013automated,koc2019empirical} have relied on annotators who inspected only the source code that warnings are reported on.
Our experiments suggest using the closed-warning heuristic, followed by manual labelling is promising -- the annotators had a strong agreement 
(Cohen's Kappa > 0.8),
while no strong agreement in manual labelling has been demonstrated in prior work.

A good benchmark requires scale and should be labelled by many annotators.
Fields such as code clone detection have created large benchmarks through community effort~\cite{roy2018benchmarks}.
This motivates the need for community effort to build a benchmark for actionable warning detection too.
As a derivative of this empirical study, we have labelled 1,300 closed warnings, usable as a starting point.

\subsection{Threats to Validity}

A possible threat to \textbf{internal validity} is the incorrect implementation of our code.
To mitigate this, we reused existing data and code whenever possible, 
including the dataset by Wang et al.~\cite{wanggolden} and Yang et al.~\cite{yangeasy,yangincremental},
and the feature extractor by Wang et al.~\cite{wanggolden}.
Our code and data are available~\cite{replication}.

Threats to \textbf{construct validity} are related to the appropriateness of the evaluation metrics.
We considered the evaluation metrics used in prior studies~\cite{wanggolden,yangeasy,yangincremental},
and also computed F1, which have been used in many classification tasks ~\cite{kim2008classifying,rahman2012recalling,zhang2020sentiment}.
F1 captures the tradeoff between Precision and Recall, and is a more appropriate measure on an imbalanced dataset.

Threats to \textbf{external validity} concern the generalizability of our findings. 
There are several threats to external validity, including the choice of projects and techniques used in our experiments.

One threat to external validity is the choice of projects studied in this paper.
We studied nine projects used in previous studies,
and we considered another set of projects that actively use Findbugs.
All considered projects  were large, mature projects.

Another threat to external validity is the choice of the approach used as an actionable warning detector.
Our analysis focuses on the use of the Golden Features SVM, 
which had the best median performance in experiments in prior studies and was the suggested model~\cite{yangeasy}.
Other approaches using different features may achieve stronger performance.

Our analysis regarding unconfirmed actionable warnings and false alarms 
also relies on human oracles (configuration/filter files written by developers, manual labelling by human annotators) that may not be perfectly accurate.
Moreover, these oracles will produce more accurate labels for warnings that are easier to label 
(e.g. shorter and well-documented code, warning types that are easier to reason about).
This may skew the distribution of labels and warnings in the datasets.
To mitigate some of the above threats, multiple annotators labeled the warnings independently, and we report the inter-rater reliability.
We achieved a strong agreement (Cohen's Kappa > 0.8).
To mitigate the threat of unmaintained Findbugs filter files,
we selected only popular projects that have filter files with at least 10 lines.

Another threat is the focus on Findbugs and Java projects.
Our analysis may not generalize to warnings of other ASATs, such as Infer~\cite{distefano2019scaling}.
This threat is mitigated as Findbugs detects a wide range of bug patterns, including bugs patterns shared by other ASATs,
and the features are not language-specific.
Moreover, we used the same dataset as prior studies~\cite{wanggolden,yangeasy,yangincremental}.
Findbugs is among the most commonly used ASATs~\cite{zampetti2017open}, having been downloaded over a million times.

\vspace{-0.1cm}
\section{Related Work}
\label{sec:related}

In Section \ref{sec:background}, we discussed the studies related to ASATs as well as the approaches that use machine learning to detect actionable warnings.
We discuss other related studies in this section.

Many studies have performed retrospectives of the state-of-the-art for various Software Engineering tasks.
Some papers~\cite{zeng2021deep,lin2021automated,liu2018neural,hellendoorn2017deep,gros2020code} study the limitations of existing tools, 
and others~\cite{lin2018sentiment,rabin2021generalizability,kang2019assessing} assess the applicability of the tools when applied to situations with a setting different from the original experiments.
Our study not only uncovers limitations of the Golden Features,  
but investigates the performance of the Golden Features under different settings (a different warning oracle in our study).

Other studies have shown the need to carefully consider data used in experiments~\cite{tu2018careful,allamanis2019adverse,zheng2015method,kalliamvakou2014promises,kochhar2014potential}.
Similar to Allamanis et al.~\cite{allamanis2019adverse}, we show that data duplication may cause overly optimistic experimental results.
Similar to Kalliamvakou et al.~\cite{kalliamvakou2014promises}, we suggest that researchers should be careful about interpreting automatically mined data.
Kochhar et al.~\cite{kochhar2014potential} investigated multiple types of bias that affect datasets used to evaluate bug localization techniques~\cite{zhou2012should,wang2014version},
and, similar to our work, find that prior experimental results were impacted by bias in the datasets.
Our work is similar to the work of Tu et al.~\cite{tu2018careful} in highlighting the problem of data leakage, where information from the future is used by a classifier and lead to overoptimistic experimental results.
Our analysis indicates that there may be delays before developers inspect static analysis warnings.
Related to this, Zheng et al.~\cite{zheng2015method} found that the status of many issues in Bugzilla may only be changed after large delays. 
These delays have implications for heuristics that are used to automatically infer labels from historical data (in our case: if a warning is actionable).

Sheppard et al.~\cite{shepperd2013data} had previously discussed data quality in a commonly used dataset for defect prediction.
While both our study as well as Sheppard et al. raise the problem of data duplication,
the duplicated instances in the dataset analyzed in this paper refer to the same warnings and labels occurring
in both training and testing dataset.
In contrast, Sheppard et al. refers to duplicated cases that occur naturally (similar features belonging to different instances, e.g. software modules).

\section{Conclusion and Future Work}
\label{sec:conclusion}

In this study, we show that the problem of detecting actionable warnings from Automatic Static Analysis Tools is far from solved.
In prior work, the strong performance of the ``Golden Features'' were 
contributed by data leakage 
and data duplication 
issues,
which were subtle and difficult to detect.

Our study highlights the need for deeper study of the warning oracle to determine ground-truth labels.
By changing the reference revision, different conclusions about performance of the Golden Features can be reached.
Furthermore, the oracle produce labels that human annotators and developers of projects using static analysis tools may not agree with.
Our experiments show that the Golden Features SVM had improved performance on cleaner data.

Our study indicates opportunities and challenges 
for future work.
It highlights the need for community effort to build a large and reliable benchmark 
and to compare newly proposed approaches with strawman baselines.
A replication package is provided at
\begin{center}
https://github.com/soarsmu/SA\_retrospective
\end{center}

\begin{acks}
  This research/project is supported by the National Research Foundation, Singapore, 
  under its Industry Alignment Fund - Pre-positioning (IAF-PP) Funding Initiative. 
  Any opinions, findings and conclusions or recommendations expressed in this material are those of the author(s) and do not reflect the views of National Research Foundation, Singapore.
\end{acks}

\balance
\bibliographystyle{ACM-Reference-Format}
\bibliography{SA_how_far}


\end{document}